# Influence of Manganese Content on Plastic Deformation Mechanisms in Polycrystalline α-Ti–Mn Alloys

G. Marković[a*], M. Fedorov[b,] M. Sokić[a], K. Frydrych[b], F. J. Dominguez-Gutierrez[b]

[a]Institute for Technology of Nuclear and Other Mineral Raw Materials, 11000 Belgrade, Serbia
[b]National Centre for Nuclear Research, NOMATEN CoE, ul. Andrzeja Sołtana 7, 05-400 Świerk, Poland

**Abstract**

Titanium alloys are widely used in aerospace, biomedical, and energy applications owing to their high specific strength, corrosion resistance, and biocompatibility. Among them, α-titanium alloys with a hexagonal close-packed (hcp) crystal structure exhibit characteristic deformation mechanisms governed by crystallographic slip and defect evolution. In this study, the influence of manganese content on the plastic deformation mechanisms of polycrystalline α-Ti–2Mn and α-Ti–4Mn (at.%) alloys is investigated using molecular dynamics simulations. Atomistic models were subjected to uniaxial loading at room temperature at a strain rate of $10^9$ s$^{-1}$. The mechanical response was evaluated through stress–strain behavior, structural evolution, dislocation nucleation and interaction, and analysis of the local deformation field. Plastic deformation in these α-Ti–Mn alloys is dominated by dislocation nucleation and their subsequent evolution within the hcp lattice. Increasing Mn content leads to higher stress levels and enhanced resistance to plastic deformation, accompanied by changes in dislocation activity and defect evolution.

## 1. Introduction

Titanium alloys combine low density with high specific strength and corrosion resistance, which makes them suitable for a wide range of aerospace, biomedical, and energy applications (Najafizadeh et al., 2024). Among them, α-titanium alloys with a hexagonal close-packed (hcp) crystal structure show characteristic deformation behavior, largely governed by crystallographic anisotropy and a limited number of active slip systems. Plastic deformation in α-Ti is mainly accommodated by dislocation slip on basal, prismatic, and pyramidal planes, while deformation twinning can be activated when slip becomes restricted (Britton et al., 2015). The balance between these mechanisms controls strength, ductility, and strain hardening, and is therefore directly linked to the evolution of crystal defects during deformation.

Alloying is commonly used to modify the mechanical response of α-titanium. Even relatively small additions of substitutional elements can affect lattice parameters, stacking-fault energies, and dislocation mobility within the hcp structure (Salloom et al., 2016). Unlike β-titanium alloys, where phase stability plays a dominant role, in α-phase systems these effects are more closely related to changes in defect energetics and slip activity. As a result, alloying influences the onset of plasticity and the way in which dislocations nucleate and evolve during deformation. Recent molecular dynamics studies have provided detailed insight into deformation mechanisms in α-Ti. In particular, they highlight the role of temperature on dislocation nucleation, mobility, and slip system activation in single-crystalline systems (Marković et al., 2025). However, the influence of alloying elements on these mechanisms, especially in polycrystalline α-Ti, remains less explored. While previous studies have extensively examined deformation mechanisms in pure titanium and single-crystal systems, the combined effect of alloying and grain structure on plastic deformation behavior in α-Ti alloys remains insufficiently understood. Among various alloying elements, manganese is of particular interest due to its ability to modify local lattice distortions and influence defect energetics, which can significantly affect dislocation behavior and plastic deformation.

Manganese is typically considered a β-stabilizing element at higher concentrations (Kim et al., 2016). However, at lower contents, where the hcp structure is retained, its role is less straightforward. In this regime, Mn acts as a substitutional solute, introducing local lattice distortions and affecting dislocation mobility and slip activity (Salloom et al., 2016). Although its influence on strength and hardness has been reported, the mechanisms by which Mn affects plastic deformation in the α-phase are still not fully clarified. In particular, there is limited atomistic insight into how manganese content influences dislocation nucleation, interaction, and strain localization. Understanding these mechanisms at the atomistic scale is essential for linking compositional effects with macroscopic mechanical behavior.

In polycrystalline α-Ti, grain boundaries additionally influence deformation by acting both as dislocation sources and as obstacles to their motion. Their interaction with solute atoms becomes particularly important under dynamic loading conditions, where plasticity is dominated by dislocation activity. However, the effect of manganese content on deformation mechanisms in polycrystalline α-Ti, especially at the atomistic level, has not been systematically investigated.

In this study, molecular dynamics simulations are employed to investigate the effect of Mn content on plastic deformation mechanisms in polycrystalline α-Ti alloys. Particular attention is given to stress–strain behavior, dislocation activity, and the spatial distribution of local deformation, providing atomistic insight into the mechanisms governing plasticity in these alloys.

## 2. Methods

- CALPHAD calculations

The phase diagram for the binary TiMn system was computed by applying the CALPHAD (CALculation of PHAse Diagrams) approach (Kaufman and Cohen, 1956; Saunders and Miodownik, 1998; Spencer, 2008). In this approach, the Gibbs energy of each relevant phase is described by physically based thermodynamic models as a function of temperature, composition, and pressure, with model parameters assessed from a combination of experimental data and first-principles calculations. The equilibrium phase assemblages were obtained by minimizing the total Gibbs energy of the system at fixed temperature and overall composition. The computations were performed using the PandaT software (Cao et al., 2009), while all thermodynamic descriptions were taken from the PanHEA-2024 thermodynamic database (Zhang and Gao, 2016), which provides assessed Gibbs energy functions for solution phases and intermetallic compounds relevant to multicomponent and high-entropy alloy systems. The phase diagram was computed over the temperature and composition ranges of interest under equilibrium conditions at ambient pressure (1 bar), assuming global thermodynamic equilibrium.

- Molecular dynamics simulations

Molecular dynamics simulations were conducted utilizing the Large-scale Atomic/Molecular Massively Parallel Simulator (LAMMPS) (Thompson et al., 2022). The main aim was to create an accurate depiction of plastic deformation, a crucial element of the mechanical reaction of materials to external pressures. An interatomic potential based on the modified embedded atom method (MEAM) was employed, specially calibrated to represent the plastic deformation characteristics of the Ti–Mn alloy (Sharifi and Wick, 2025). The potential was obtained by an extensive fitting process that included experimental observations, density functional theory (DFT) calculations, and thermodynamic modeling. Consequently, it precisely replicates essential crystallographic characteristics, such as β-phase stability, elastic constants, and stacking fault energy (Sharifi and Wick, 2025) rendering it very appropriate for the current investigation.

1. Samples preparation

HCP Ti-2Mn and Ti-4Mn (at.%), as well as pure HCP Ti samples for tensile test simulations were generated using Atomsk (Hirel, 2015). A unit cell of pure HCP titanium was first defined with lattice constants a=0.2951 nm and c=0.4683 nm. Next, Vornoi tessellation method was used to seed the simulation box with this unit cell and generate eight grains with random orientations with the help of polycrystal mode of Atomsk. The dimensions of the simulation box are 40x15x15.2 nm and it has fully periodic boundary conditions, containing 516374 atoms. The elongated shape of the sample in $x$ direction is necessary for the tensile test simulations. Five different random polycrystals were generated, and their composition altered accordingly, in order to isolate the effect of the chemical composition in the polycrystals with identical grain structure (Dominguez-Gutierrez et al., 2026). For Ti-Mn alloys, manganese atoms were then randomly substituted into the titanium lattice to achieve concentrations of 2 and 4 at.% Mn. Although the absolute grain sizes in molecular dynamics simulations are smaller than those in experimental specimens due to computational limitations, the statistical nature of MD simulations gives insight into local deformation mechanisms that are scalable to experimental conditions. Energy minimization was carried out using the conjugate gradient algorithm, ensuring that the system reached its lowest-energy configuration. The minimization criteria required that both the energy difference between consecutive iterations and the relative energy magnitude remain below $10^{-5}$ eV, while the total force on each atom was constrained to $\leq 10^{-8}$ eV/Å (Cichocki et al., 2025). The equilibrated samples were subsequently relaxed at room temperature for 2 ns in an isothermal–isobaric (NPT) ensemble. Temperature and pressure control were achieved through integration of the Nosé–Hoover equations, employing damping parameters of 2 fs for the thermostat and 5 ps for the barostat, under an external pressure of 0 GPa (Cichocki et al., 2025, Domínguez-Gutiérrez et al., 2024).

2. Tensile test setup

Tensile test simulations were performed with a strain rate of $10^9$ s$^{-1}$. In each computational step, the constituent particles within the atomistic ensemble were remapped according to the instantaneous dimensions. This ensured displacement-controlled straining of the cuboidal simulation cell along the straining direction (the *x*-axis), while a barostat was applied in the remaining two directions (*z* and *y*) to maintain constant pressure in those directions (Cichocki et al., 2025, Domínguez-Gutiérrez et al., 2024). The applied stress was calculated directly from the MD simulations by considering the number of atoms in the system and their atomic volumes. The uniaxial strain imposed on the system was calculated using the following relation:

$$\varepsilon = \frac{l_0 - l_x(t)}{l_0} \tag{1}$$

where $l_0$ represents the initial cell length along the x-axis before strain was applied, and $l_x(t)$ is the instantaneous cell length along the x-axis at time *t*. The time step for the simulation was set to 2 fs to ensure the accuracy and stability of the results throughout the deformation process.

In order to investigate the localization of the stress inside the samples, the general von Mises stress have been calculated for each atom as (Hill, 1998):

$$\sigma_v = \sqrt{\frac{1}{2}[(\sigma_{11} - \sigma_{22})^2 + (\sigma_{22} - \sigma_{33})^2 + (\sigma_{33} - \sigma_{11})^2] + 3(\sigma_{12}^2 + \sigma_{23}^2 + \sigma_{31}^2)} \tag{2}$$

where $\sigma_{ij}$ are the elements of the Cauchy stress tensor.

Structural and defect analysis was performed using the OVITO software package (Stukowski et al., 2010). The dislocation density was calculated using the Dislocation Extraction Algorithm (DXA) implemented in OVITO (Stukowski and Albe, 2010), which identifies dislocation lines based on the local atomic structure and corresponding Burgers vectors.

## 3. Results

The phase diagram calculated for the binary Ti–Mn system is shown in Fig. 1. The system exhibits a complex phase stability, particularly in the Mn-rich regime, with multiple intermetallic phases and two eutectic reactions at higher Mn contents. The liquidus temperature reaches its maximum for pure Ti (approximately 1670 °C) and decreases with increasing Mn content. Showing that low Mn concentrations is relevant for the present study, a body-centered cubic (BCC, β-Ti) phase is predicted at temperatures above ~550 °C. In a narrow compositional range, a coexistence of hexagonal close-packed (HCP, α-Ti) and BCC phases is also observed. Upon decreasing temperature, equilibrium conditions suggest the formation of α-Ti together with secondary phases.

However, it should be noted that the phase diagram represents equilibrium conditions, whereas molecular dynamics simulations inherently describe non-equilibrium atomic configurations. In the present study, Ti–2Mn and Ti–4Mn at.% alloys are therefore modeled as single-phase HCP systems, since the simulations focus on deformation behavior at room temperature and do not explicitly account for phase transformations. This assumption is consistent with typical atomistic approaches and enables a direct analysis of defect-mediated plasticity within the α-phase.

Example of the polycrystalline simulation box used in tensile tests is presented in Fig. 2(a), and the following analysis, where not presented for all samples, is based on this sample with varying concentration of Mn. At the start of the simulations, grains bounaries occupy approximately 18% of the sample.

The stress–strain curves in Fig. 2(b) show the expected sequence of elastic response, yielding, and plastic deformation. A clear trend is observed: increasing Mn content leads to higher resistance to deformation over the entire strain range. The onset of plasticity is also shifted, indicating that the presence of Mn affects dislocation nucleation.

The absolute stress levels differ from experimental values, which is expected for molecular dynamics simulations performed at high strain rates. In addition, the interatomic potential used here does not fully reproduce elastic constants and defect energetics. For this reason, the discussion focuses on relative trends rather than direct quantitative agreement.

The strengthening behavior can be explained by solid solution effects. The substitution of Ti atoms by Mn introduces local lattice distortions in the hcp structure due to atomic size mismatch, generating stress fields that interact with dislocations and increase resistance to their motion. In α-Ti, plastic deformation is mainly governed by ⟨a⟩-type dislocations (Barkia et. al., 2017) particularly on prismatic planes, whose nucleation and mobility depend strongly on the local atomic environment. Alloying can also modify the relative stability of slip configurations through changes in generalized stacking fault energy. Variations in stacking fault energies influence dislocation dissociation and the activation of specific slip systems, thereby affecting both dislocation nucleation and mobility. This effect is particularly important in hcp structures, where the limited number of slip systems makes plastic deformation more sensitive to such energetic changes.

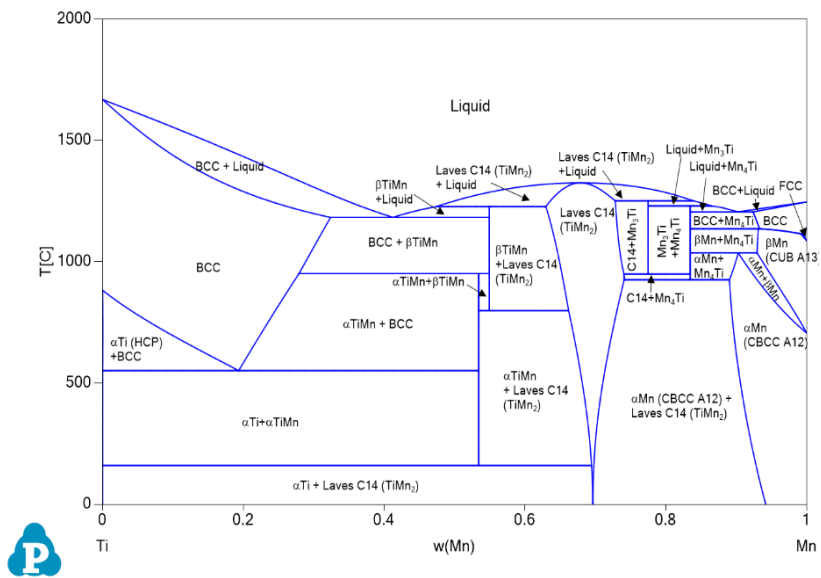

Fig. 1. Phase diagram of Ti-Mn system, calculated with PanHEA-2024 thermodynamic database in PandaT at 0 pressure (Chen et al, 2002).

As a result, the combined effect of local lattice distortion and modified defect energetics leads to reduced dislocation mobility and increased resistance to plastic deformation. The interaction between solute atoms and dislocation cores further increases the energy barriers for slip, which is consistent with the observed strengthening trend.

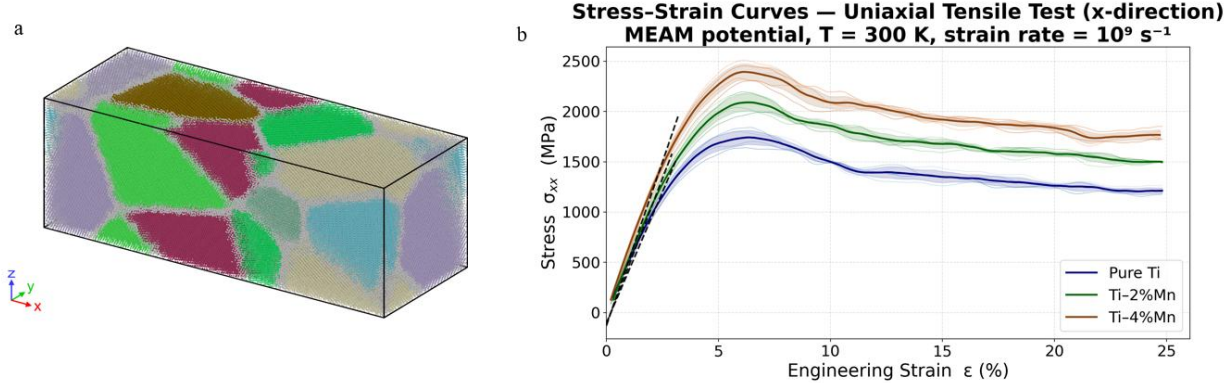

Fig. 2. (a) Example of the initial simulation cell for Ti, Ti–2Mn, and Ti–4Mn used in MD simulations, color represents different grains; (b) Stress–strain curves for pure Ti, Ti–2Mn and Ti–4Mn.

Fig. 3 shows the grains and stacking faults at the final step of the tensile test simulation (a-c), as well as evolution of the fraction of FCC-ordered atoms within the grains (d–f), and the dislocation density (g–i) as a function of applied strain for pure Ti, Ti–2Mn, and Ti–4Mn.

As can be seen in this figure, both the FCC percentage (d-f) and the dislocation density (g–i) start forming after yielding, reflecting the activation of plastic deformation mechanisms. While the overall trend is similar for all compositions, subtle differences indicate that Mn influences dislocation nucleation and interaction processes. It should be noted, that due to the polycrystalline nature of the investigated systems, there is no strong correlation between FCC percentage and dislocation density. One of the differences in the behavior of pure Ti and the Ti-Mn alloys is the thickness of the grain boundaries: in Ti, the atomic percentage of grain boundaries increases much faster than in Ti-Mn alloys, and at the end of simulations reaches the value of 58%, compared to 46% in Ti-2Mn and 39% in Ti-4Mn at.%.

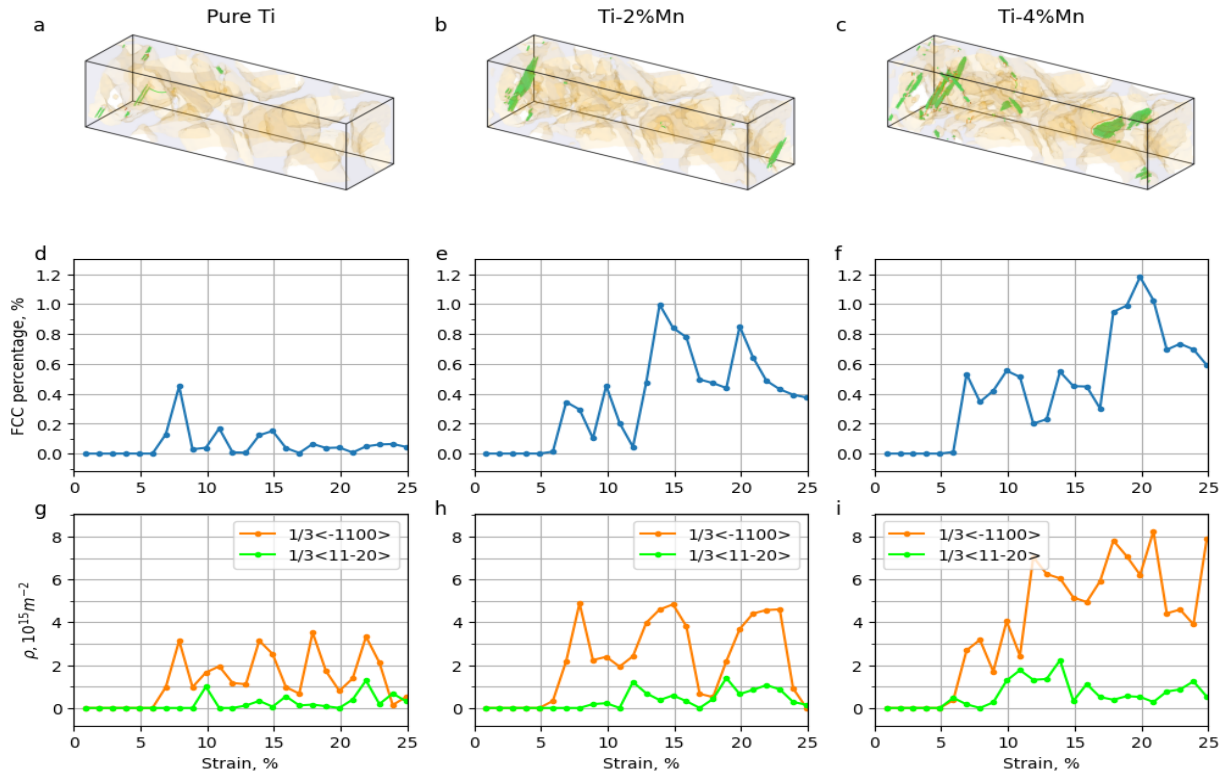

Fig. 3. Visualization of the final atomic configuration of the simulation cell after the strain test (a-c), percentage of the FCC-ordered atoms inside the grains (d-f), dislocation density ρ in $10^{15}m^{-2}$ as a function of applied strain (g-i). In (a-c), the grains are shown as transparent surfaces, and green atoms are the FCC atoms forming stacking faults.

Apart from the grain boundary growth, a different mechanism of deformation is occurring inside the grains. When strain is applied, stacking faults are formed (which in HCP material are planes ordered as FCC), limited inside the grain by the 1/3<1-100>-type edge dislocations. Alternatively, they are limited by the grain boundaries. The most noticeable example of this mechanism can be seen in Fig. 3(h) for Ti-2%Mn sample, where the increase of FCC percentage correlates with the increase in the 1/3<1-100>-type dislocation density between 12% and 15% of strain. Under further application of strain, three different options are possible.

1) Two 1/3<1-100>-type dislocations surrounding a stacking fault form one 1/3<11-20>-type dislocation, that is terminated on the grain boundaries, following the balance of energies, e.g. 1/3[-1010]- 1/3[1-100]=1/3[-2110]. This leads to a decrease in the percentage of FCC atoms.

2) Stacking fault is absorbed into the grain boundary, leading to a decrease in the density of 1/3<11-20>-type dislocations.

3) Stacking faults grow until they fully terminate on the grain boundaries and either stabilize or split the grain in two.

Stacking faults form in every sample, when the stress is applied. Most noticeable difference is that in alloys with Mn, their number and size are increased. Feature noticeable with increasing Mn concentration to 4% compared to 2% is the formation of systems containing 3-5 parallel stacking faults forming inside large grains.

The spatial distribution of the von Mises stress for pure Ti, Ti–2Mn, and Ti–4Mn is presented in Fig. 4 (a), (b) and (c), respectively. The values were computed in OVITO using the general form of the von Mises stress which is a scalar measure of local deformation field. Grain boundaries are left uncolored, as stress inside them varies significantly from atom to atom, diluting the visual clarity. The decrease of the grain boundary volume is clearly seen in Fig. 4 (d), when comparing the area of grain boundaries in pure Ti, Ti-2%Mn and Ti-4%Mn.

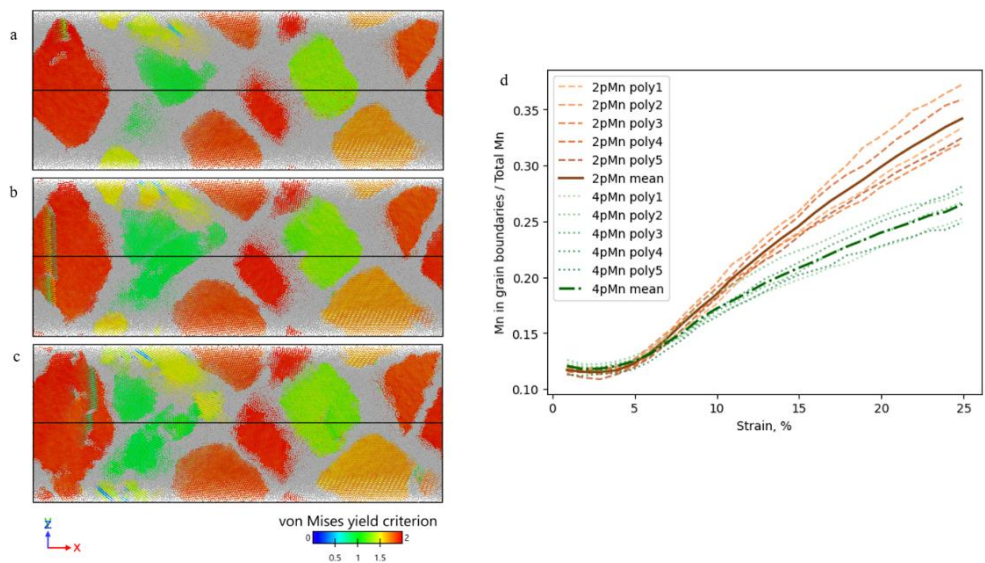

Fig. 4. Atomic von Mises strain distribution at the final stage of deformation for (a) pure Ti; (b) Ti–2Mn; and (c) Ti–4Mn. In (d), the fraction of all Mn atoms found in grain boundaries is presented.

The other important feature present in von Mises stress maps are the stacking faults, visible in the left part of each subfigure of Fig. 4. As indicated by the color, the local stress of the stacking fault is significantly lower compared to the stress of the grain in which the respective stacking faults are found.

Lastly, different stages of grain splitting can be seen when comparing the green features near the leftmost large grain, in Fig. 4(b) the green grain is intact, while in Fig. 4(a) and Fig. 4(c) it is split into multiple smaller grains.

Figure 4(d) shows the fractions of all Mn atoms that is found inside the grain boundaries as a function of strain (while the rest of the Mn atoms are inside the grains). It can be seen that both for Ti-2%Mn and Ti-4%Mn, the fraction of atoms inside the grain boundaries increases with strain, indicating the migration of Mn atoms from the grains into grain boundaries, with faster rate for Ti-2%Mn as compared to Ti-4%Mn.

## 4. Conclusions

This study investigates the influence of manganese on the plastic deformation behavior of polycrystalline α-Ti using molecular dynamics simulations. A clear trend is observed: increasing Mn content leads to higher resistance to deformation, as reflected in the stress–strain response. At the atomic scale, plastic deformation is governed by dislocation nucleation and their subsequent interactions. The addition of Mn modifies the local atomic environment, hindering dislocation mobility and contributing to the observed strengthening. Changes are also evident in the spatial distribution of deformation. While deformation in pure Ti is more uniformly distributed, the addition of Mn leads to a more localized response, indicating a transition toward more heterogeneous plastic flow. Overall, these results show that even small additions of Mn can significantly affect defect activity and deformation behavior in α-Ti, highlighting the role of alloying in controlling plasticity at the atomic scale.

**Acknowledgement**

Research was funded through the European Union Horizon 2020 research and innovation program under Grant Agreement No. 857470 and the initiative of the Ministry of Science and Higher Education "Support for the activities of Centers of Excellence established in Poland under the Horizon 2020 program" under Agreement No. MEiN/2023/DIR/3795. The Ministry of Science, Technological Development, and Innovation of the Republic of Serbia, grant No. 451-03-136/2025-03/200023. We gratefully acknowledge Polish high-performance computing infrastructure PLGrid (HPC Center: ACK Cyfronet AGH) for providing computer facilities and support within computational Grant No. PLG/2024/017084.

# References

Najafizadeh, M., Yazdi, S., Bozorg, M., Ghasempour-Mouziraji, M., Hosseinzadeh, M., Zarrabian, M., Cavaliere, P., 2024. Classification and applications of titanium and its alloys: A review. Journal of Alloys and Compounds Communications 3, 100019.

Britton, T.B., Dunne, F.P.E., Wilkinson, A.J., 2015. On the mechanistic basis of deformation at the microscale in hexagonal close-packed metals. Proceedings of the Royal Society A: Mathematical, Physical and Engineering Sciences 471, 20140881.

Salloom, R., Banerjee, R., Srinivasan, S.G., 2016. Effect of β-stabilizer elements on stacking fault energies and ductility of α-titanium using first-principles calculations. Journal of Applied Physics 120, 17.

**Marković, G., Domínguez-Gutiérrez, F.J., 2026. Thermally activated plasticity in single-crystal titanium: A molecular dynamics study of nanoscale deformation. Modelling and Simulation in Materials Science and Engineering 34, 015018.**

Kim, J.W., Hwang, M.J., Han, M.K., Kim, Y.G., Song, H.J., Park, Y.J., 2016. Effect of manganese on the microstructure, mechanical properties and corrosion behavior of titanium alloys. Materials Chemistry and Physics 180, 341–348.

Kaufman, L., Cohen, M., 1956. The martensitic transformation in the iron–nickel system. JOM 8, 1393–1401.

Saunders, N., Miodownik, A.P. (Eds.), 1998. CALPHAD (Calculation of Phase Diagrams): A Comprehensive Guide. Elsevier

Spencer, P.J., 2008. A brief history of CALPHAD. Calphad 32, 1–8.

Cao, W., Chen, S.L., Zhang, F., Wu, K., Yang, Y., Chang, Y.A., Schmid-Fetzer, R., Oates, W.A., 2009. PANDAT software with PanEngine, PanOptimizer and PanPrecipitation for multi-component phase diagram calculation and materials property simulation. Calphad 33, 328–342.

Zhang, C., Gao, M.C., 2016. CALPHAD modeling of high-entropy alloys. In: High-Entropy Alloys: Fundamentals and Applications. Springer, Cham, pp. 399–444.

Thompson, A.P., Aktulga, H.M., Berger, R., Bolintineanu, D.S., Brown, W.M., Crozier, P.S., in't Veld, P.J., Kohlmeyer, A., Moore, S.G., Nguyen, T.D., Shan, R., Stevens, M.J., Tranchida, J., Trott, C., Plimpton, S.J., 2022. LAMMPS – a flexible simulation tool for particle-based materials modeling at the atomic, meso, and continuum scales. Computer Physics Communications 271, 108171.

Sharifi, H., Wick, C.D., 2025. Developing interatomic potentials for complex concentrated alloys of Cu, Ti, Ni, Cr, Co, Al, Fe, and Mn. Computational Materials Science 248, 113595.

Hirel, P., 2015. Atomsk: A tool for manipulating and converting atomic data files. Computer Physics Communications 197, 212–219.

Cichocki, K., Dominguez-Gutierrez, F., Wyszkowska, E., Kurpaska, L., Muszka, K., 2025. Evaluating compression and nanoindentation in FCC nickel: A methodology for interatomic potential selection. Archives of Mechanics 77, 437–459.

Domínguez-Gutiérrez, F.J., Olejarz, A., Landeiro Dos Reis, M., Wyszkowska, E., Kalita, D., Huo, W.Y., Jozwik, I., Kurpaska, L., Papanikolaou, S., Alava, M.J., Muszka, K., 2024. Atomistic-level analysis of nanoindentation-induced plasticity in arc-melted NiFeCrCo alloys: The role of stacking faults. Journal of Applied Physics 135, 18.

**Hill, R., 1998. The mathematical theory of plasticity. Oxford University Press, Oxford.**

**Dominguez-Gutierrez, F.J., Frelek-Kozak, M., Markovic, G., Strozyk, M. A., Daramola, A., Traversier, M., Fraczkiewicz, A., Zaborowska, A., Khvan, T., Jozwik, I., Kurpaska, L., 2026. Phys. Rev. Materials 9, 123607**

Stukowski, A., 2010. Visualization and analysis of atomistic simulation data with OVITO. Modelling and Simulation in Materials Science and Engineering 18, 015012.

**Stukowski, A., Albe, K., 2010. Extracting dislocations and non-dislocation crystal defects from atomistic simulation data. Modelling and Simulation in Materials Science and Engineering 18, 085001.**

**Chen, S. L., Daniel, S., Zhang, F., Chang, Y.A., Yan, X. Y., Xie, F. Y., Schmid-Fetzer, R., Oates, W. A., 2002. Calphad 26, 175**

Barkia B, Couzinié JP, Lartigue-Korinek S, Guillot I, Doquet V. In situ TEM observations of dislocation dynamics in α titanium: Effect of the oxygen content. Materials Science and Engineering: A. 2017 Aug 4;703:331-9.